\documentclass[aps,prb,showpacs,floatfix,amsmath,amssymb,
reprint,superscriptaddress]{revtex4-1}
\usepackage{color}
\usepackage{graphicx}
\usepackage{natbib}
\usepackage{epsfig}
\usepackage{setspace}
\usepackage{amsmath,amssymb}
\usepackage{verbatim}
\usepackage{bibentry}

\def\etal{{\em{et al. }}}

\begin{document}
\title{A local theory for Mott-Anderson localization}

\author{Sudeshna Sen} 
\affiliation{Theoretical Sciences Unit, 
Jawaharlal Nehru Centre for Advanced 
Scientific Research, Bangalore-560064, India}
\author{Hanna Terletska}
\affiliation{Department of Physics, University of Michigan, Ann Arbor, Michigan 48109, USA}
\author{Juana Moreno}
\affiliation{Department of Physics \& Astronomy, Louisiana State University, Baton Rouge, Louisiana 70803, USA}
\affiliation{Center for Computation \& Technology, Louisiana State University, Baton Rouge, Louisiana 70803, USA}
\author{N. S. Vidhyadhiraja} \email{raja@jncasr.ac.in}
\affiliation{Theoretical Sciences Unit, 
Jawaharlal Nehru Centre for Advanced 
Scientific Research, Bangalore-560064, India}
\author{Mark Jarrell}\email{jarrellphysics@gmail.com}
\affiliation{Department of Physics \& Astronomy, Louisiana State University, Baton Rouge, Louisiana 70803, USA}
\affiliation{Center for Computation \& Technology, Louisiana State University, Baton Rouge, Louisiana 70803, USA}
\begin{abstract}
The paramagnetic metallic phase of the Anderson-Hubbard model 
(AHM) is investigated using a non-perturbative local moment approach within the framework of dynamical mean field theory with a typical medium. Our focus is on the breakdown of the metallic phase near the metal-insulators transition as seen in the single-particle spectra, scattering rates and the associated distribution of Kondo scales.  We demonstrate the emergence of a universal, underlying low energy scale, $T_K^{peak}$. This lies 
close to the peak of the distribution of Kondo scales obtained within the 
metallic phase of the paramagnetic AHM. Spectral dynamics for energies, $\omega\lesssim T_K^{peak}$ display Fermi liquid universality crossing over to an incoherent universal dynamics for $\omega\gg T_K^{peak}$ in the scaling regime.  
Such universal dynamics indicate that within a local theory the low to moderately low energy physics is governed by an effective, {\it disorder renormalised} Kondo screening.
\end{abstract}

\maketitle

\section{Introduction}
Disorder is ubiquitous in real materials, strongly influencing their properties \cite{mirlin_review_2008,mackinnon_1993,fifty_years}. 
Another aspect of several condensed matter systems 
like the heavy fermions or transition metal oxides is the presence of strong electron-electron interactions \cite{Phys_today_DMFT, DMFT_RMP}. 
In particular, Coulomb correlations and disorder may individually drive a system towards a metal-insulator 
transition. While the Anderson metal-insulator transition \cite{Anderson} is caused by quenched disorder, the Mott-Hubbard metal-insulator transition emerges from strong 
Coulomb repulsion \cite{DMFT_RMP}. The simultaneous presence of disorder ($W$) and interaction ($U$) effects is known 
to influence material properties in subtle ways. Over the last few decades, several experimental works on a range of systems
\cite{mirlin_review_2008, kravchenko_review_2004, Miranda1997_review, 
perovskite_maiti, perovskite_kim, perovskite_athena, cold_atom1, cold_atom2, 
cold_atom3} have highlighted the importance of the interplay of disorder and interactions.
 The early theoretical studies of such systems \cite{TVR_Lee_RMP} mainly focused on the weak disorder limit, perturbing around Fermi liquid theory. It is now known that the subtle interplay of disorder and 
interactions may lead to non-Fermi liquid like responses in the thermodynamic quantities, as observed in several experiments \cite{Miranda1997_review}; 
therefore, one requires a non-perturbative framework that can deal with interactions and disorder on an equal footing.

One of the most intriguing aspects of strongly correlated electron systems is the appearance of low-energy scales \cite{Raja_dyn_sca_pam, Si_Steglich_QCP_review, Bulla2001}. 
For metals with strong electronic correlations a frequently observed scenario is 
the presence of long-lived quasi-particles representing a coherent Fermi liquid picture at the lowest temperatures ($T$) and energy scales. A universal low energy scale, $T^*$, lies at the heart of all 
strongly correlated electron systems that manifests in the universal transport properties of these 
materials \cite{Bulla2001}. Over the past few decades, the dynamical mean field theory (DMFT) has stood out as a very successful theoretical framework for understanding several aspects of the 
low energy physics of strongly correlated electron systems \cite{DMFT_RMP}. Since, in DMFT, 
any lattice-fermion model reduces to a {\it local} quantum impurity model, the involvement of Kondo physics is inevitable. Thus, universal behavior due to the emergence of a universal 
low energy scale is generally attributed to the underlying Kondo effect. For example in the DMFT picture of the metallic phase of the Hubbard model \cite{DMFT_RMP,Logan_mott_ins}, the Kondo effect 
leads to full quenching of the electron spin degrees of freedom resulting in a non-degenerate Fermi liquid ground state characterized by a low energy Fermi liquid scale. In the vicinity of the Mott transition, the strongly correlated metal is therefore characterized by a low-energy 
scale corresponding to the coherence temperature of a Fermi liquid \cite{DMFT_RMP,Logan_mott_ins}. 

The emergence of a single low energy/temperature scale may not be restricted to clean 
strongly-correlated electron systems, but has also been predicted in the context of 
diluted two-dimensional electron gases (2DEGs) \cite{2DEGs_vlad}. Phenomenological theories, based on experimental observations in 2DEGs \cite{kravchenko_review_2004,Kravchenko2,Kravchenko3}, established a similarity between the metal-insulator transitions in such disordered systems and the conventional Mott-Hubbard metal-insulator transition. Studies in these
directions are important for understanding the true driving force behind metal-insulator transitions observed in disordered interacting systems. A finite temperature study of the effects of disorder on the 
non-zero temperature Mott transition \cite{disorder_Tneq0} also revealed the prevalence of a single parameter scaling of the distribution of quasi-particle weights in the vicinity of a 
disordered Mott transition. 

A natural question that 
follows from these studies is whether such scaling with respect to a 
single low energy scale also manifests in the dynamics of the microscopic 
quantities like the single particle spectra or the disorder averaged 
scattering rate in a disordered interacting system at zero 
temperature. And if such universal dynamics exists, then how 
general is this observation across the $W-U$ phase diagram? The origin 
and evolution of such low energy scales with respect to 
$W$ and/or $U$ would then reflect upon the driving force behind 
the localization of the electrons. 

The understanding of the behavior of the low energy scales in a strongly correlated system thus 
stand out as a key prerequisite irrespective of the presence or absence of disorder. 
Several theoretical frameworks have attempted to understand the interplay of disorder and strong correlations \cite{AA,Punnoose2005,finkelstein}. However, the study of emergent low energy scales 
in the simultaneous presence of disorder and electron-electron interactions require non-perturbative 
frameworks. Studies using the framework of the DMFT have provided several insights in these 
directions. A computationally inexpensive approach involves the framework of DMFT 
and utilization of the `typical' density of states (TDoS, $\rho_{typ}$) \cite{vlad_nikolic,schubert} for self-consistently obtaining the effectively local hybridizing medium, $\Gamma(\omega)$\cite{Byczuk2005, Aguiar_2006, Aguiar2009, 50_years_vollhardt, aguiar_2013}, in which the single impurities are embedded. This construction of the DMFT bath utilizing the $\rho_{typ}(\omega)$ is known as the TMT-DMFT framework 
\cite{50_years_vollhardt}. 
The TDoS is most 
appropriately approximated as the geometric average of the 
local density of states (LDoS), 
$\langle\rho(\omega)\rangle_{geom}=\rho_{typ}(\omega)=
\exp\langle\ln\rho_i(\omega)\rangle$ with 
$\rho_i(\omega)$ being the local density of states. Another way of representing this is
$\rho_{typ}=\exp \int dV_i P(V_i) \ln\rho_i(\omega)$, where 
$V_i$ represents the bare random potential and 
$P(V_i)$, the probability distribution followed by 
these bare site energies. While, $\rho_{typ}(0)$ is critical 
at the Anderson 
transition \cite{Vollhardt_review,first_TMT_Vlad2003}, the average 
density of states (ADoS) given by 
$\rho_{arith}(\omega)=\int dV_i P(V_i) \rho_i(\omega)$ 
is not critical. The TDoS behaves like an order parameter for 
the metal-insulator transition originating in the metallic 
phase; in the insulating phase it is trivially zero at 
all frequencies. Thus, in principle the TMT framework is 
designed to capture the physics of the 
metal-insulator transition approaching from the metallic phase.

The TMT-DMFT method was first applied to the Anderson-Hubbard model by Byczuk \etal \cite{Byczuk2005} who explored the $W-U$ paramagnetic phase diagram using numerical renormalization group (NRG) as the impurity solver. 
Three distinct phases were identified namely, 
a correlated metallic phase, a Mott insulating phase and an Anderson 
insulating phase. Additionally, a coexistent regime of the 
metal-Mott insulating phase was reported. 
The Mott and Anderson insulator phases were 
found to be continuously connected. The characterization of these phases were 
based on the behavior of the band center of the TDoS ($\rho_{typ}(0)$) and 
the ADoS ($\rho_{arith}(0)$). The metallic phase 
featured a non-zero $\rho_{typ}(0)$ and $\rho_{arith}(0)$. 
For weak to 
moderate $W$, a sharp transition from the metallic to a gapped insulating 
phase was observed where both $\rho_{typ}(0)=0$ and 
$\rho_{arith}(0)=0$. This metal-insulator transition 
was similar in characteristics with the 
conventional single band Hubbard model and hence 
this insulating phase was termed as the Mott 
insulator. Moreover, the density of states in this phase featured prominent 
Hubbard subbands. Additionally a metal-Mott insulator 
coexistence regime similar to the p-h symmetric 
single-band Hubbard model 
was identified in the $W$-$U$ plane that terminated at a single $W$. The Anderson insulator was characterized as a phase that 
featured $\rho_{typ}(0)=0$ and $\rho_{arith}(0)\ne0$. 
Additionally, the Hubbard bands were broad and diffused.

Although, the NRG is highly efficient in capturing the Kondo effect, the distribution of Kondo scales, a natural occurrence in interacting disordered systems, was not explored in 
Ref.~\onlinecite{Byczuk2005}. Thus the role of the local Kondo scales could not be deduced from the above calculation. Such a direction was however explored using slave-boson mean-field theory calculations \cite{Aguiar2009}, highlighting the role of the local quasi-particle weights, $Z_i$, that may also serve as an 
order parameter for the localization physics in the 
Anderson-Hubbard model. 
Close to the disorder driven metal-insulator transition 
at $U/W<1$ a {\it two fluid} picture was proposed. Through TMT-DMFT calculations they proposed a spatially inhomogeneous picture where in certain regions there existed Mott fluid droplets with $Z_i\to 0$ and at other regions $Z_i\to 1$ representing Anderson localized particles. Irrespective of the spatially inhomogeneous 
picture, one would expect a metal-insulator transition 
to occur at a critical disorder strength, 
$W_c$, when the $U$ is fixed, and this would coincide 
with the vanishing of the impurity hybridization obtained 
from the TDoS. A conventional Mott-like picture was proposed to prevail for the $U$ driven metal-insulator transition 
at sufficiently small disorder strengths. A similar line of reasoning based on the 
behavior of the impurity hybridization would lead us to 
expect that the Mott upper critical interaction, 
$U_{c2}$,  would coincide 
with the vanishing of the impurity hybridization obtained 
from the TDoS.

However, it is also well known that slave-boson based solvers 
fail to account for inelastic scattering and thus fail to predict the correct lineshape of spectral functions and scattering rates \cite{senechal2006theoretical,SB_artefact1}. 
Moreover, the physics at low energies, may be highly affected by the physics at higher  
energy scales. 
Thus, in order to have a precise understanding of the 
spectral/dynamical 
properties in a correlated system, we require all energy scales and 
interaction strengths, from weak to strong coupling, to be handled 
within a unified theoretical framework.
In this work we revisit the metallic phase of the Anderson-Hubbard model using the local moment approach (LMA) \cite{Eastwood1998} as an impurity solver within the TMT-DMFT. The LMA has been successfully applied for several impurity
\cite{Eastwood1998, LMA_pseudogap_AIM1, LMA_NRG_soft_gap_AIM} and 
lattice models \cite{Eastwood1997,  LMA_KIs, Raja_dyn_sca_pam} 
(within DMFT). 
The LMA is known to capture the Kondo 
effect correctly while also capturing the correct lineshape of the spectral 
functions. With this set up we look into the evolution of the 
distribution of Kondo scales as a function of $W$ and $U$. Additionally, we explore the scattering dynamics within 
the current non-perturbative local framework, and identify universal dynamics 
and scaling similar to the clean interacting scenario. It should be noted that all the calculations presented 
in this work pertain to the metallic phase and an exploration 
of the insulating phases is beyond the scope of the current 
work. These results are therefore relevant in the context of 
the breakdown of the 
metallic phase towards Mott or Anderson localization.  

\section{Model}
The Anderson-Hubbard model is considered as a paradigmatic model for looking into the 
interplay of strong electron interactions and disorder. It is given by, 
\begin{align}
  \hat{\mathcal{H}}=
  -&\sum_{\langle ij\rangle, \sigma}t_{ij}\left(c_{i\sigma}^{\dagger}c_{j\sigma}+ \mathrm{H.c.}\right)
  +\sum_{i, \sigma}(V_i-\mu)\hat{n}_{i\sigma}\nonumber \\  
              +&U\sum_i\hat{n}_{i\uparrow}\hat{n}_{i\downarrow},
  \label{eq:AHM}
\end{align} where, $c_{i\sigma}^{\dagger}$ ($c_{i\sigma}$) is the fermionic creation  
(annihilation) operator for an electron with spin $\sigma$ at site $i$, 
and $\hat{n}_{i\sigma}=c_{i\sigma}^{\dagger}c_{i\sigma}$, 
$t_{ij}$ is the nearest neighbour site to site hopping 
amplitude considered 
to be constant in this work, 
$U$ is the onsite Coulomb interaction energy. The lattice is represented by a three-dimensional cubic DoS with full bandwidth, $D=3$. The random local potential $V_i$ follows a "box" distribution $P(V_i)$ such that $P(V_i)=\frac{1}{2W}\Theta(W-\vert V_i\vert)$, where $\Theta(x)$ is the Heaviside step function. 
A global particle-hole symmetry is imposed by $\mu=U/2$.  
At $W=0$, this model reduces to the 
particle-hole (p-h) symmetric single-band Hubbard model, which displays a first order Mott transition at zero temperature, $T=0$, as a function of 
$U$. On approaching this transition from the Fermi liquid (FL) side, the Kondo scale, $T_K^0$ 
vanishes at a critical point, $U_{c2}$, marking the transition to the Mott insulating state.  On approaching from the Mott insulating side, the Mott gap 
vanishes at a critical point, $U_{c1}$, where, $U_{c1}<U_{c2}$.  
This scenario in the $W=0$ case motivates us to look at the regimes, $U<U_{c2}$ and $U>U_{c2}$ distinctly. 
For a three-dimensional simple cubic DoS, within the LMA, 
we have found out that the Mott MIT occurs at $U_{c2}/D\sim0.8$ 
which corresponds to $U_{c2}\approx 2.3$, the 
bandwidth ($D$) being equal to 3. This result compares well with the value 
predicted by NRG calculations ($\sim1.1D$) \cite{bonvca2009van}.

For treating non-zero disorder in the presence of interactions 
(Eq.~\eqref{eq:AHM}) we employ the TMT-DMFT framework where we map the disordered lattice model on to an ensemble of 
single impurity Anderson models, each embedded in a 
self-consistently determined effective medium, $\Gamma(\omega)$, which is obtained from $\rho_{typ}(\omega)$, as described 
in Appendix~\ref{app2}. The 
reader is also referred to several previous works 
\cite{Byczuk2005, Aguiar_2006, Aguiar2009, aguiar_2013, Vollhardt_review} for 
the details of the formalism.  In Appendix~\ref{app1} and 
~\ref{app2} we also describe the implementation of the LMA 
within the TMT-DMFT framework. 
We typically solve for $\sim10^5$ disorder realizations each of which involves the calculation of the local interaction self-energy, $\Sigma_i(\omega)$.  

\section{Results and discussions}
In the absence of interactions, Eq.~\eqref{eq:AHM} reduces to the 
Anderson model of non-interacting electrons \cite{Anderson}. Here, 
the metal to insulator transition is not characterized by 
the vanishing of the DoS. Instead, the 
hybridization paths get canceled accompanied by weak  
localization of the wave functions due to coherent 
backscattering from impurities or exponential 
localization of the wave functions in deep-trapped states
\cite{Bulka1985, Bulla2001, mackinnon_1993}. 
As a result, the electrons occupying such exponentially 
localized states are confined to limited regions in the 
space and cannot contribute to the transport. As the disorder 
potential, $W$, is increased, more and more regions in space 
become exponentially localized and the  
system undergoes a metal-insulator transition as a function 
of $W$. At the Anderson localization transition 
the average DoS given by, 
$\rho_{arith}(\omega)=\int dV_i P(V_i) \rho_i(\omega)$, 
with $\rho_i(\omega)$ being the LDoS, 
is not critical. However, the geometrical mean of the  
LDoS, $\rho_{typ}=\exp \int dV_i P(V_i) \ln\rho_i(\omega)$, better approximates the critical nature 
of the Anderson localization transition. 
The local TMT framework adopted here reproduces some of the expected features of the Anderson localization transition, but underestimates the critical disorder strength, $W_c$ \cite{first_TMT_Vlad2003}. Although by construction 
the local TMT framework is able to describe qualitatively the 
effects of strong localization due to disorder, all non-local
coherent backscattering effects are lost. The 
localization mechanism {\it explicitly} contained within the 
TMT is essentially the physics due to deep-trapped states where 
the states initially above and below the bare band-edge  
become localized in deeply trapped states
\cite{mackinnon_1993, Bulka1985, Bulka1987}. This effect is 
subsequently fed back into the hybridizing medium so that the 
band center also localizes. Within TMT, the band edge of 
$\rho_{typ}(\omega)$ then monotonically moves towards the 
band center such that at the critical disorder strength 
even states at the band center are exponentially localized. 

Perturbative studies on the weakly interacting disordered 
electron gas dates back to the seminal work of Altshuler and 
Aronov \cite{AA}. 
Later extensions include the two-loop large-$N$ 
reormalization group analysis of Punnoose and Finkelstein 
\cite{Punnoose2005}, that could describe a metal-insulator 
transition in a two-dimensional electron gas. However, in disordered 
interacting systems there also exists a number of 
relevant phenomena that are beyond the reach of perturbative methods. For example, the work  by Milanovi{\'c}, Sachdev and Bhatt \cite{Milov1989} and 
later by Bhatt and Fisher \cite{Bhatt1992} showed the importance of disorder in describing 
the instability of a disordered, interacting Fermi liquid 
towards the formation of local moments. 
The treatment of 
interactions within the non-pertubative 
framework of DMFT \cite{DMFT_RMP} can naturally incorporate 
the tendency towards the formation of local moments \cite{Aguiar2009}. 
\begin{figure}[tp!]
    \centerline{\includegraphics[clip=,scale=0.5]
    {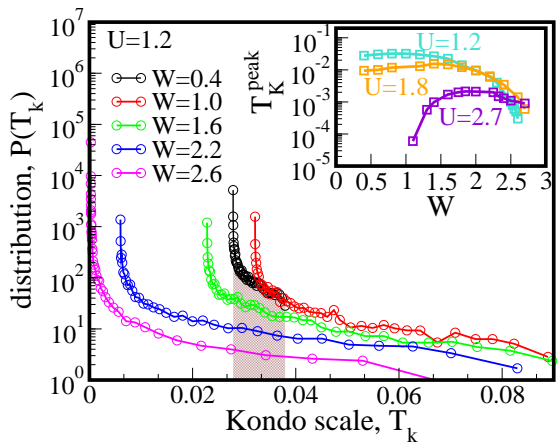}}
    \caption{{\bf Distribution of Kondo scales:}  
    In the main panel, the evolution of the 
    $T_K$ distribution as a function of $W$ for $U=1.2$  
    is shown on a linear-log scale.
  The distributions are peaked and (sharply) bounded from below 
  by $T_K^{peak}$, the scale associated with the respective  
  particle-hole symmetric limit of the effective impurity problem 
  embedded in the {\it typical} medium. The shaded region 
  highlights the narrow range of $T_K$'s spanned by small 
  $W$ in contrast 
  to the higher $T_K$ long tails spanned by larger $W$'s. 
  (inset) $T_K^{peak}$ is plotted as a function of $W$ for 
  $U=1.2,\,1.8,\,2.7$. While, $U=1.2,\,1.8$ correspond to $U<U_{c2}$, 
  $U=2.7>U_{c2}$. 
  Hence, in the $W=0$ limit $U=1.2$ and 
  $U=1.8$ correspond to Fermi liquids with 
  $T_K^0=T_K^{peak}\approx0.025$ and $0.007$, respectively. 
  On the other hand, $U=2.7$ corresponds to a Mott 
  insulator with $T_K^0=T_K^{peak}=0$.}
    \label{fig:TK_distribn}
  \end{figure}
In this work, we revisit the paramagnetic phase of the AHM 
and try to elucidate the mechanism that could lead to the 
formation of such local moments in a disordered, interacting 
system. In particular, we    
look into the single particle quantities across a broad range of 
$U$ and $W$ parameters, putting particular 
emphasis on the scattering rate and the evolution of the distribution of Kondo scales with respect to $U$ and $W$.

\subsection{Distribution of Kondo scales:}
It is well known that the metallic DoS of the particle-hole 
symmetric single-band Hubbard model 
exhibits a three peak structure, with a 
well defined Abrikosov-Suhl resonance centered around the Fermi energy, that signifies 
the low-energy quasiparticle coherence present in the system, symptomatic 
of an underlying coherence scale, $T_K^0$ 
\cite{DMFT_RMP}. The full width at half maximum of this resonance is one measure of the low energy Kondo coherence scale, $T_K^0$, present in the Fermi liquid. The local quasi-particle weight, $Z=\left(1-\frac{\partial\mathrm{Re}\Sigma(\omega)}{\partial\omega}\right)^{-1}$ provides another measurement of this energy scale. Above this coherence scale, physical properties are dominated by incoherent electron-electron scattering effects and Fermi liquid theory loses its validity  although, recent state-of-art DMFT calculations indicate a {\it resilient quasi-particle} regime before 
the system crosses over to a bad metal regime \cite{deng2013bad}.
In the presence of disorder the translational invariance is broken, so the screening of the local moments 
by mobile electrons should be spatially non-uniform. While some sites may be strongly hybridized 
with the local medium, others may be weakly 
hybridized. For sites that are 
weakly hybridized with the local surroundings charge fluctuations
are suppressed, thus representing a reduced 
screening in comparison to the sites that strongly hybridize with the 
surrounding medium. Therefore, in a 
strongly correlated disordered system, the  coherence scale should in 
principle be viewed as a random position dependent quantity with an associated 
distribution. 

Within the TMT-DMFT implementation we solve an ensemble of impurity problems embedded in an 
effective disorder averaged medium. We use the LMA as our impurity solver.
The LMA is designed to capture the dynamical spin flip scattering processes encountered by an $\uparrow/\downarrow$ spin occupied impurity. These processes lie at the heart of the physics 
associated with the Kondo effect \cite{LMA_SIAM1}, and their energy scale 
is on the order of 
the Kondo scale. The LMA can capture such extremely low energy scales 
efficiently. Within the LMA, a measure of the Kondo scale is provided by the 
position of the resonance in the transverse spin polarization propagator 
\cite{LMA_SIAM1}. 
We therefore end up with a self-consistently determined distribution of such spin-flip scattering energy scales that represent the energies associated 
with the Kondo screening of the impurities by the disorder averaged effective non-interacting host.
 \begin{figure}[tp!]
\centerline{\includegraphics
   [clip=,scale=0.5]
   {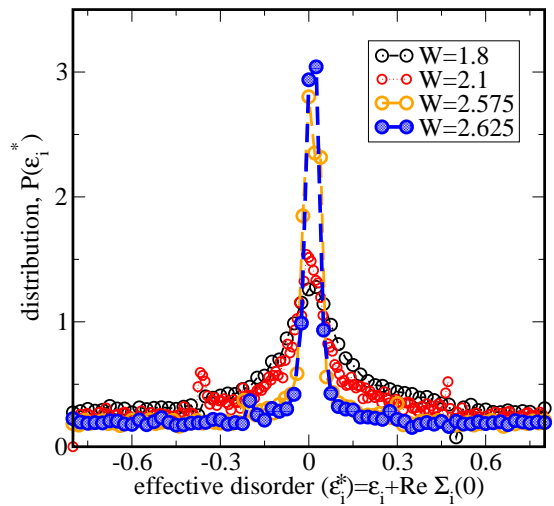}}
  \caption{{\bf Distribution of the renormalized site energy}, 
$\epsilon_i^*=\epsilon_i+\mathrm{Re}\Sigma_i(0)$, with 
$\epsilon_i=V_i-U/2$, plotted for $U=1.2$ and 
  $W=1.8,\,2.1,\,2.575,\,2.625$. 
A pronounced weight is observed around $\epsilon_i^*=0$ that represents the 
particle-hole symmetric limit. The initially broad peak becomes narrower and 
grows in intensity as $W$ is increased. Such an evolution of $\epsilon_i^*$ 
as a function of increasing $W$ indicates that a majority of sites tend to 
attain a $T_K$ close to that of the particle-hole symmetric limit. This can be 
correlated with the skewed nature of 
$P(T_K)$ in Fig.~\ref{fig:TK_distribn} as $W$ is increased. 
Note that the entire range is not shown.}
  \label{fig:eff_potn}
\end{figure}

In Fig.~\ref{fig:TK_distribn} we show the
distributions of Kondo scales, $T_K$, for various disorder strengths, $W$, 
at a fixed $U=1.2$. The local nature of the framework renders 
the distributions to be peaked and bounded from below 
(also observed in earlier works at non-zero temperature  \cite{disorder_Tneq0} and square lattice \cite{Vlad_griffiths_phase}).  
This peak, 
$T_K^{peak}$, is associated with the 
particle-hole (p-h) symmetric limit of the 
effective impurity problem embedded in the disorder averaged medium that is also p-h symmetric and is identical for all such single-impurity sites. Due to the local nature of the solution, the effective Kondo screening experienced by any impurity moment could 
thus be only dependent on the $V_i$ of the respective impurity.  Therefore sites which are at or close to the p-h 
symmetric limit will experience the least Kondo screening
and hence will have the lowest $T_K$.
The shaded region in Fig.~\ref{fig:TK_distribn} 
demonstrates the narrow area under the curves corresponding to the low 
disorder limit of $W=0.4$ for $U=1.2$, in contrast to the long tails in the distributions corresponding to the higher values of $W$. 
\begin{figure}[h!]
    \centerline{\includegraphics[clip=,scale=0.3]
    {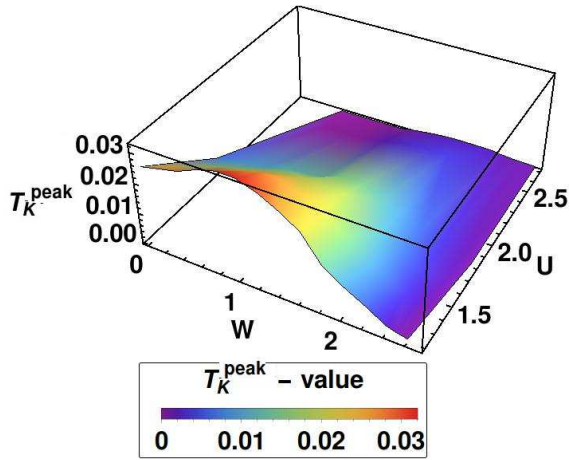}}
\caption{A schematic of the surface formed by the $T_K^{peak}$ 
as a function of $U$ and $W$.}
\label{fig:schematic1}
\end{figure}
The initial effect of 
increasing $W$ is to screen the effects of $U$ even at the lowest energy 
scales, such that $T_K^{peak}$ is pushed to higher values; 
subsequently, with increasing $W$, $T_K^{peak}$ decreases monotonically, signifying the onset of 
disorder induced scattering cooperating with interaction driven 
scattering in the low frequency region, and tending to localize the system. 

In the inset of Fig.\ref{fig:TK_distribn} we plot 
$T_K^{peak}$ as a function of $W$ for different interaction strengths, $U$. 
When $W=0$, the systems with $U=1.2$ and $1.8$ are Fermi liquids with Kondo scales
$T_K^0=T_K^{peak}\approx0.025$ and $0.007$, respectively.
For $U=2.7$, the system is 
a Mott insulator with $T_K^0=T_K^{peak}=0$.
\begin{figure*}[tp!]
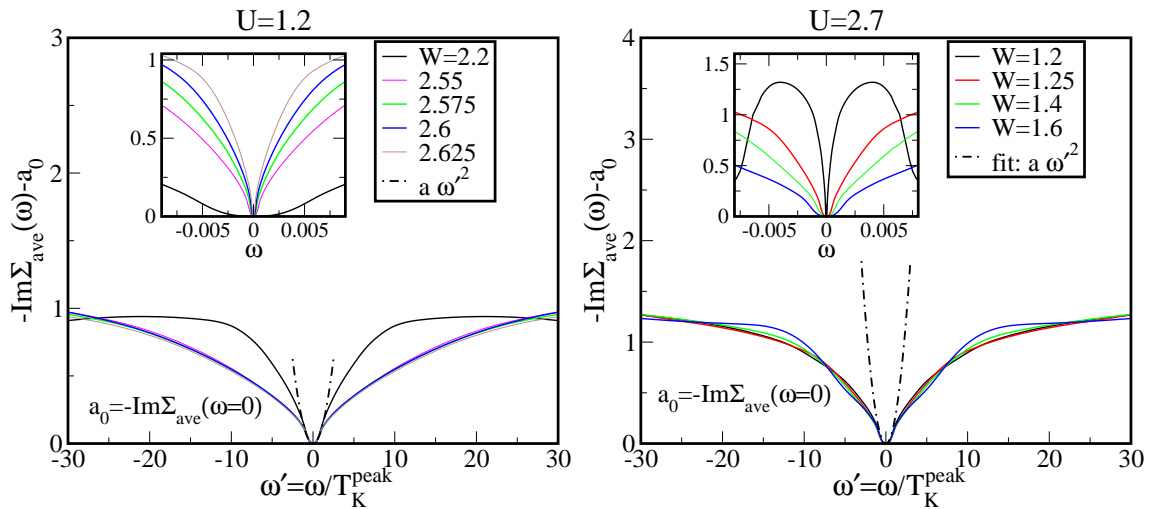

\centerline{\includegraphics
   [clip=,scale=0.5]
   {Figure4a.eps}
{\includegraphics
   [clip=,scale=0.5]
   {Figure4b.eps}}}
\caption{{\bf Universal scaling of -Im$\Sigma_{ave}(\omega)$:}
(Left main panel) The 
$-\mathrm{Im}\Sigma_{ave}(\omega)$ with the static part 
($a_0=-\mathrm{Im} \Sigma_{ave}(0)$) 
subtracted is plotted for $U=1.2$, on an energy scale, 
$\omega^\prime$, with 
the bare frequency, $\omega$ 
rescaled by $T_K^{peak}$ that has been obtained from the 
respective $P(T_K)$ plots.
(Left inset) The same sets of data are plotted on a 
bare energy scale, $\omega$.
(Right main panel) The 
$-\mathrm{Im}\Sigma_{ave}(\omega)$ with the static part 
($a_0=-\mathrm{Im} \Sigma_{ave}(0)$) 
subtracted is plotted for $U=2.7$ on the rescaled 
energy scale, $\omega^\prime$. 
(Right inset) The same sets of data are plotted on a 
bare energy scale.}
 \label{fig:rescale_SE_U1.2_2.7}
\end{figure*}
As shown in the inset of Fig.\ref{fig:TK_distribn}, 
for $U=2.7$ $(U>U_{c2})$, the $T_K^{peak}$ evolves from being zero at low $W\ll U$, and then at $W=W_{c1}$ a non-zero $T_K^{peak}$ emerges that  subsequently increases with increasing disorder signifying a regime where disorder screens the effects of strong interactions. This initial screening of electron-electron interactions due to disorder is true even for smaller $U=1.2$ or $U=1.8$ ($U<U_{c2}$) as discussed earlier. Subsequently, $T_K^{peak}\to 0$ as $W$ is increased, an observation that holds true for both $U=1.2$ and $U=1.8$.

Particular insight about 
the respective behavior of $P(T_K)$ may be obtained by 
looking at the evolution of the {\it effective} site potential 
energy ($\epsilon_i^*$) as a function of increasing $W$. 
In Fig.~\ref{fig:eff_potn} we show 
the distribution  
of the disorder renormalised site energy, $P(\epsilon_i^*)$, 
for $W=1.8,\; 2.1,\;2.575,\;2.625$ at $U=1.2$, 
with $\epsilon_i^{*}=V_i-U/2+\mathrm{Re}\Sigma_i(0)$. 
The distribution is marked by a peak 
around $\epsilon_i^{*}=0$, indicating that a majority of sites tend to attain a disorder renormalized site potential energy close to 
the p-h symmetric limit.
This explains why the $T_K^{peak}$ is determined by the Kondo scales corresponding to the sites that are at or close to half-filling. This peak is initially broad for a relatively low $W$ ($W=1.8$) and becomes sharper as the $W$ is increased. This shows that as the disorder is increased more number of sites experience a reduced 
Kondo screening. In other words, for stronger $W$'s, the 
distribution of Kondo scales become more and more skewed such 
that even sites that are quite far away from half filling may experience a 
reduced Kondo screening resulting in a Kondo energy 
scale close to that corresponding to the p-h symmetric limit.

It is to be noted that such a behavior of $P(\epsilon_i^*)$ has 
been shown to exist close to an {\it interaction} driven transition
at a fixed $W$ \cite{aguiar_2003, Andrade20093167}, where 
such a behavior of $P(\epsilon_i^*)$ has been dubbed as {\it 
perfect disorder screening}. In other words similar observations 
were shown to be prevalent close to the metal-Mott insulator 
phase boundary of the symmetric, paramagnetic AHM 
\cite{aguiar_2003, Andrade20093167}. 
In this work, we show that this behavior of $P(\epsilon_i^*)$ and 
$P(T_K)$ is generic to a broader parameter regime. These 
observations are  
not just restricted to $U$ driven metal-insulator transition 
at low $W$, 
but applies to $W$ driven transitions also even if the bare 
interaction strength is small. The physical picture underlying 
the above observation is the following:
for 
strong disorder potential, as we approach a disorder driven 
metal-insulator transition, the $\Gamma_{typ}(0)$ 
between any site and its host becomes sufficiently small such that the ratio, 
$U/\pi\mathrm{Im}(\Gamma_{typ}(0))\gg 1$, and these  sites with $\epsilon_i^*\sim 0$ experience stronger interaction effects, pushing $T_K^{peak}$ towards zero, even though the bare 
interaction strength is small. 
In Fig.~\ref{fig:schematic1} we summarize the above analysis by representing the surface of $T_K^{peak}$ scales as a function of both $U$ and $W$.
Since $T_K^{peak}$ represents the most probable value of the underlying Kondo scale, we now ask the question whether this can be related to the scattering dynamics of the system close to the metal-insulator transitions observed in the AHM. 
In the following section we therefore explore the 
imaginary part of the disorder averaged self-energy, -Im$\Sigma_{ave}(\omega)$.

\subsection{Scattering dynamics}

In a strongly correlated system the imaginary part of the interaction  
self-energy, -Im$\Sigma(\omega)$, relates to the scattering rate. 
Thus the -Im$\Sigma(\omega)$ is a mirror of the underlying 
scattering dynamics present. In a disordered 
interacting system we need to look at the average self-energy, 
-Im$\Sigma_{ave}(\omega)$, obtained from the arithmetically 
averaged Green's function, 
$\langle G(\omega)\rangle_{arith}$, where $\langle\dots\rangle_{arith}$ denotes arithmetic averaging 
with respect to $P(V_i)$. It is this average quantity 
that represents the 
physical Green's function of the system. The quantity, 
$\langle G(\omega)\rangle_{arith}$ may be obtained 
from the Hilbert transform of $\rho_{arith}(\omega)$, 
given by, $\langle G(\omega)\rangle_{arith}=
\int\frac{\rho_{arith}(\omega^\prime)
d\omega^\prime}{\omega-\omega^\prime}$. Accordingly, 
the average self-energy, that 
represents the scattering dynamics, is obtained from 
the Dyson's 
equation given by
$\Sigma_{ave}(\omega)
=\mathcal{G}(\omega)^{-1}-\langle G(\omega)\rangle_{arith}^{-1}$. 
The host Green's function $\mathcal{G}(\omega)$ embodies the {\it typical} nature of the disorder-averaged medium. 
\begin{figure}[t!]
    \centerline{\includegraphics[clip=,scale=0.225]
    {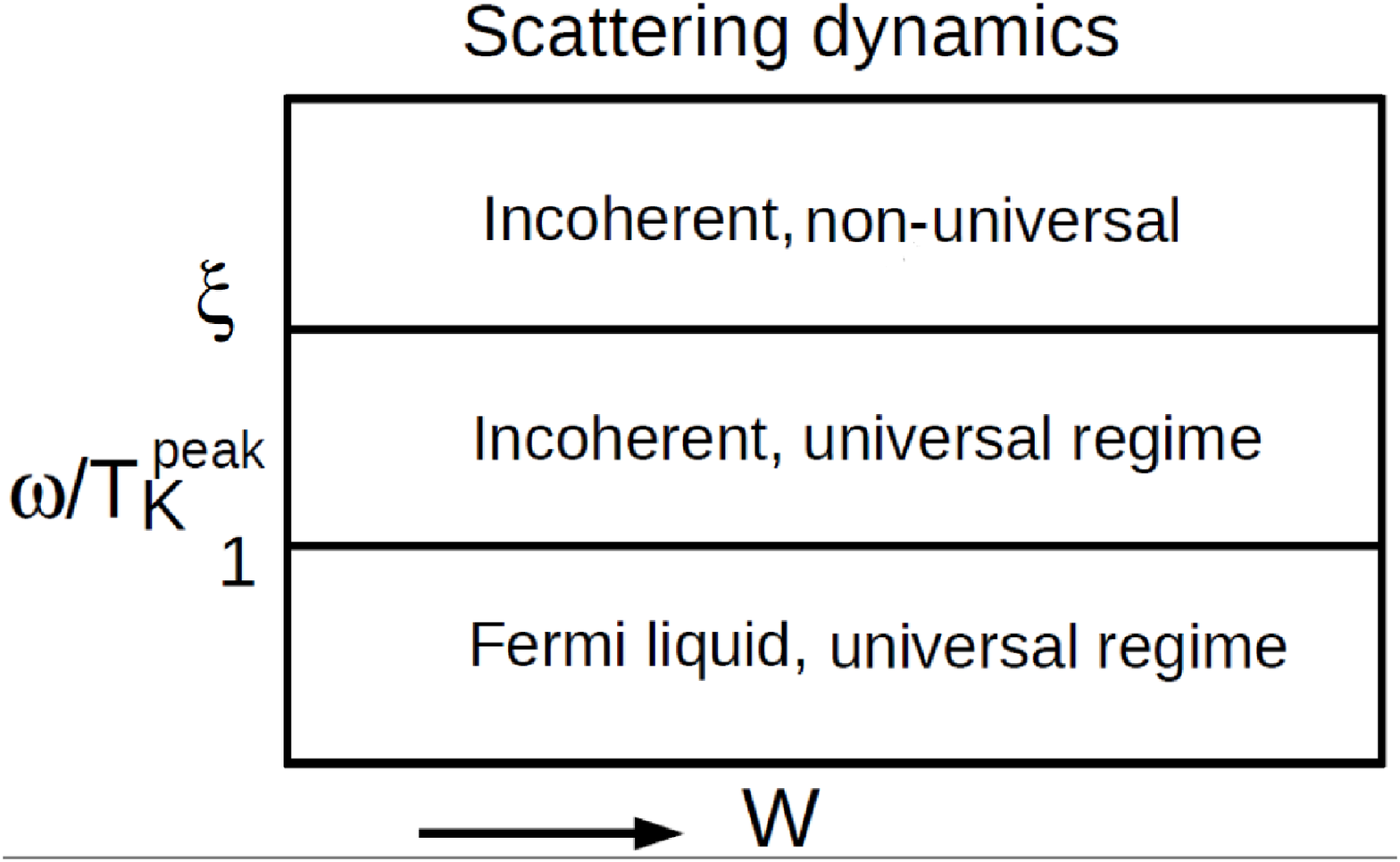}}
\caption{A schematic demonstrating the self-energy scattering dynamics corresponding to different energy regimes as a 
function of disorder, $W$. 
It should be noted that the 
the quantity $\xi$ separating the `incoherent universal' 
and the `incoherent non-universal' regimes is just 
symptomatic of the proximity to the metal-insulator 
transition. In strong 
coupling as $W\to W_c$, where $W_c$ is the critical 
disorder strength when metal-insulator transition is 
approached from the 
metallic side, $\xi\to\infty$.}, in the scaling regime.
\label{fig:schematic2}
\end{figure}

In a clean Fermi liquid, 
the $-\mathrm{Im} \Sigma_{ave}(\omega)=
-\mathrm{Im} \Sigma(\omega)\sim |\omega|^2$ for 
$\omega<T_K^0$ (where $T_K^0$ is the lattice-coherence scale or the 
Kondo scale in a clean lattice). With this background we look at 
the low frequency region of -Im$\Sigma_{ave}(\omega)$. 
In Fig.~\ref{fig:rescale_SE_U1.2_2.7}, we subtract the static contribution 
of the impurity scattering, 
$a_0=-\mathrm{Im}\Sigma_{ave}(0)$ and plot the quantity, 
-Im$\Sigma_{ave}(\omega)-a_0$ for 
various disorder strengths at fixed interaction strengths, 
$U=1.2$ (left panel) $U=2.7$ (right panel). The parameters 
presented are close to the disorder 
driven metal-insulator transition boundary. 
In the main panels we plot this quantity 
on a frequency rescaled, 
$\omega^{\prime}=\omega/T_K^{peak}$ axis, relating $T_K^{peak}$ 
to the inverse scattering rate of the particles. The insets 
to these figures show the low frequency part of the self-energy spectrum, -Im$\Sigma_{ave}(\omega)-a_0$, 
on an absolute scale, i.e. vs. $\omega$. In either 
case ($U=1.2$ or $U=2.7$) of Fig.~\ref{fig:rescale_SE_U1.2_2.7} 
the self-energy spectrum for various $W$'s look quite 
distinct on the bare energy scale being dependent on the 
disorder strength, $W$. These plots also reflect upon the 
diminution 
of the {\it effective} Kondo scale as $W$ is increased, thus 
relating to Fig.~\ref{fig:TK_distribn}. In contrast to the 
insets of Fig.~\ref{fig:rescale_SE_U1.2_2.7}, the main panels of 
Fig.~\ref{fig:rescale_SE_U1.2_2.7} illustrate the 
self-energy spectrum on a rescaled axis, with the rescaled 
frequency, $\omega^\prime=\omega/T_K^{peak}$. It is also observed 
that the low energy spectral dynamics of 
-Im$\Sigma_{ave}(\omega)-a_0 \sim \omega^{\prime 2}$ for 
$\omega^\prime<1$. At higher energy scales, a clear departure 
from $\sim\omega^{\prime 2}$ is evident as anticipated.  
A universal scaling collapse of the single-particle 
self-energy, with respect to $T_K^{peak}$ is observed reminiscent of the conventional 
correlated lattice scenario \cite{Logan_mott_ins, Zitko_Shastry}.  
The clear collapse due to this 
rescaling suggests that, within a local theory, 
even in presence of a random potential, 
an energy scale 
$\sim T_K^{peak}$ serves as a Fermi liquid scale, 
just as in the clean case. 
Moreover, as seen from the main panel of Fig.~\ref{fig:rescale_SE_U1.2_2.7}, although the 
coherent Fermi liquid scattering regime is restricted 
for $\omega^\prime\leq1$, a universal scattering 
dynamics is significantly observed until much higher 
energy scales. 
In other words, this 
signifies that within a local theory for 
interacting disordered systems, the 
quasi-particle excitations are in fact determined by a 
disorder renormalized single impurity Kondo scale, 
$T_K^{peak}$. Let us now comment on the parameter 
regime where this collapse is most significantly observed. 
The scaling collapse for $U=1.2$ holds true for higher disorders 
and very close to the transition where the $T_K^{peak}$ itself is 
exponentially small. Note that the values of 
$W=2.575, \;2.6, \;2.625$ in 
Fig.~\ref{fig:rescale_SE_U1.2_2.7} correspond to very small scales in 
Fig.~\ref{fig:TK_distribn}. So, in 
Fig.~\ref{fig:rescale_SE_U1.2_2.7}(left panel) representing $U=1.2$, 
the $W$'s represent values close to the metal to a disorder driven Mott-Anderson 
insulator transition. 
A similar scenario is observed for $U=2.7$, as shown in the left panel of 
Fig.~\ref{fig:rescale_SE_U1.2_2.7}. 
If we now locate $W=1.2, \;1.25, \;1.4$ in Fig.~\ref{fig:TK_distribn} 
(violet curve), these values would approximately correspond to 
$T_K^{peak}\sim 0.00009, \; 0.0005, \;0.001$ respectively, and would 
thus represent $W$'s close to a metal-insulator transition resembling a clean 
Mott transition. 
Note that according to Ref.~\onlinecite{Byczuk2005}, at  
the critical disorder strength where this metal-insulator transition 
would occur the $\rho_{typ}(0)$, $\rho_{arith}(0)$ 
would vanish simultaneously, `on the spot'. In Ref.~\onlinecite{Byczuk2005},  
the insulating phase resulting from this transition was termed as the disordered 
Mott insulator phase. In accordance with 
the observations of Ref.~\onlinecite{Aguiar2009}, we also speculate that the 
$T_K^{peak}$ would continuously vanish to zero at the critical $W$.
So, as observed in 
Fig.~\ref{fig:rescale_SE_U1.2_2.7}, universal scaling, until 
$\omega\gg T_K^{peak}$ is observed from $W=1.2$ and $W=1.25$, representing 
parameters very close to the {\it disordered Mott transition}. 
Note that 
since we could not reach such low energy scales for the metal to Mott-Anderson insulator 
transition at higher $W$ for $U=2.7$, 
demonstrating such a scenario in this regime 
was beyond the scope of the current work.  

\begin{figure*}[t!]
    \centerline{\includegraphics[clip=,scale=0.475]
    {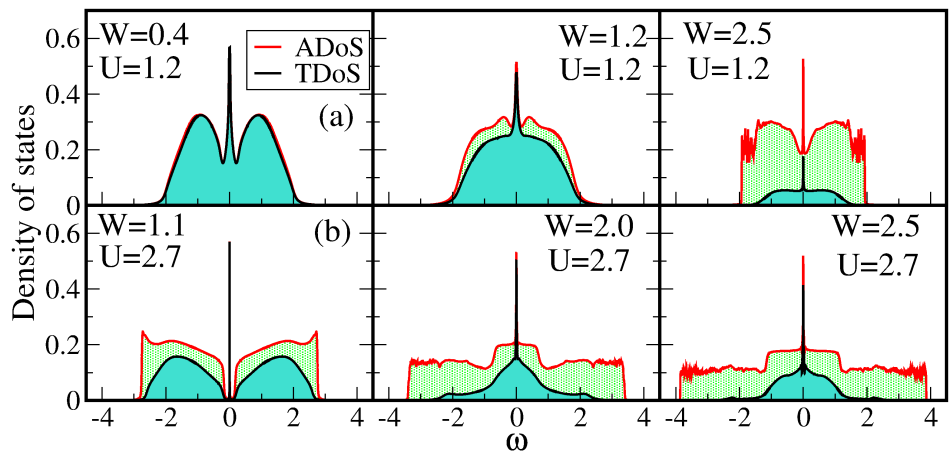}}
    \caption{{\bf Evolution of the density of states:} The arithmetic mean 
    of the density of states (DoS) also known as 
    the average DoS (ADoS) (solid red line, 
    green shaded region) and the geometric mean of the DoS 
    also known as the typical DoS (TDoS) (solid black line, turquoise shaded region) at various 
    $W$ for (a) $U=1.2$, (b) $U=2.7$. Similar to the $U=0$ scenario, when the disorder, $W$, 
is small, both the ADoS and the TDoS produce almost the 
same density of states and as $W$ increases the TDoS 
gets suppressed over all energy scales.}
    \label{fig:DOS}
  \end{figure*}
We note in passing, that such a universal scaling collapse 
scenario could already be anticipated from Fig.~\ref{fig:eff_potn} 
where we demonstrated the evolution of the distribution of 
the renormalised site energies, $\epsilon_i^*$ as a function 
of increasing disorder. As $W$ is increased in presence of 
a fixed $U$, the pronounced tendency of an appreciable 
number of sites to acquire a renormalised 
site potential, $\epsilon_i^*=0$, already 
reflect upon the possible emergence of a universal low energy 
scale close to the disorder driven metal-insulator 
transition. This in turn manifests as a universal 
scaling collapse in the spectral dynamics of 
-Im$\Sigma_{ave}(\omega)-a_0$. 
This {\it renormalized} single-particle dynamics 
is summarized in Fig.~\ref{fig:schematic2} as a schematic. 
Such universal physics determined by a single energy 
scale, even in the presence of strong disorder, suggests that the local 
effect of disorder is to only renormalize the onsite 
interaction between the electrons, such that the underlying 
low energy quasiparticle excitations are still determined 
by Fermi liquid dynamics, similar to a conventional 
Mott transition scenario. This is possibly a consequence of 
the underlying scattering mechanism due to deep-trapped 
states prevalent within a local theory and its 
resulting feedback to the low energy sector of 
the (local) hybridizing medium. 
It is worth mentioning that a similar  
universal scaling scenario of the single particle 
density of states was hinted at in an earlier study 
by Aguiar \etal in Ref.~\onlinecite{Aguiar_2006}. 
They considered an ensemble of single impurity Anderson models 
embedded in a model bath. The model bath was manually chosen, 
and the {\it typical} nature of the 
hybridization function was parametrized 
in order to mimic a disorder driven metal-insulator transition.   
In this work we elucidate and demonstrate a universal scaling 
picture of scattering dynamics within a self-consistent 
scheme, where the {\it typical} medium is 
determined self-consistently 
and depends on the amount of disorder present.

\subsection{Density of states} 
The arithmetically averaged DoS (ADoS) is defined as, 
$\rho_{arith}(\omega)=\int\,P(V_i)\rho_i(\omega)\,dV_i$, where, 
$\rho_i(\omega)$ is the local DoS (LDoS). The typical DoS is obtained via geometric averaging of the LDoS and is 
defined as, $\rho_{typ}(\omega)
=\exp\int\,P(V_i)\ln\rho_i(\omega)\,dV_i$. As mentioned earlier, 
$P(V_i)$ represents the distribution followed by the random 
site potential energies, that in this paper is chosen to be a 
box distribution.
In Fig.~\ref{fig:DOS}(a), we plot the arithmetically averaged DoS (ADoS) and 
the typical DoS (TDoS) for various $W=0.4,\;1.2,\;2.5$ at a fixed 
interaction strength 
$U=1.2 < U_{c2}$. In 
agreement with the $U=0$ scenario, when the disorder, $W$, 
is small, both the ADoS and the TDoS produce almost the 
same density of states. With increasing disorder, the TDoS 
gets suppressed over all energy scales (note that this is 
not spectral weight transfer, as the TDoS is not 
normalized).  As seen from Fig.~\ref{fig:DOS}(a) there exists remnants of the 
$W=0$ limit Kondo 
resonance centered around $\omega=0$. With increasing disorder, this 
resonance initially broadens but then progressively narrows down.
In Fig.~\ref{fig:DOS}(b), we plot the same as above but for 
$U=2.7 > U_{c2}$ that represents a Mott 
insulator in the $W=0$ limit of the p-h symmetric AHM. 
The introduction of randomness 
allows for local charge fluctuations that in turn leads to 
delocalization of the otherwise localized moments, beyond 
a certain critical disorder strength, $W_{c1}$. This picture 
is in agreement with the NRG calculations of 
Ref.~\onlinecite{Byczuk2005}. 
This naturally manifests as the emergence of a finite 
density of states at the Fermi level ($\omega=0$). In other words, a 
sharp Kondo resonance reappears in the middle of a prominent gap, with the 
inclusion of a finite amount of disorder, $W_{c1}$; 
this gap reminds us of the 
Mott insulating gap in the $W=0$ limit. Based on the spectral fingerprints, 
we may speculate the following: if we start from 
$W=1.1$ at $U=2.7$ and decrease $W$ we should expect a 
metal-insulator transition at $W_{c1}$. This transition 
is similar to the Mott metal-insulator transition 
obtained in the conventional single band Hubbard model. 
This should be reflected as the narrowing of $(P(T_K)$ 
(not shown here for $U=2.7$) and 
an associated decrease in $T_K^{peak}$ as $W$ is pushed 
towards $W_{c1}$. The latter is illustrated in the inset 
of Fig.~\ref{fig:TK_distribn}. 
For both Fig.~\ref{fig:DOS}(a) and 
Fig.~\ref{fig:DOS}(b), the high energy Hubbard bands broaden and acquire 
reduced spectral intensities. This broadening that is also manifested in the
 self-consistently determined hybridization function (not shown here) highlight the fact 
that presence of disorder introduces additional scattering pathways. In 
the context of DMFT, this increases the rate at which these high energy 
electrons hop off from the impurity site into the embedding host, thus 
reducing its lifetime and hence broadening the spectra at such energy scales.

\begin{figure}[htb!]
  \centerline{\includegraphics
   [clip=,scale=0.5]
   {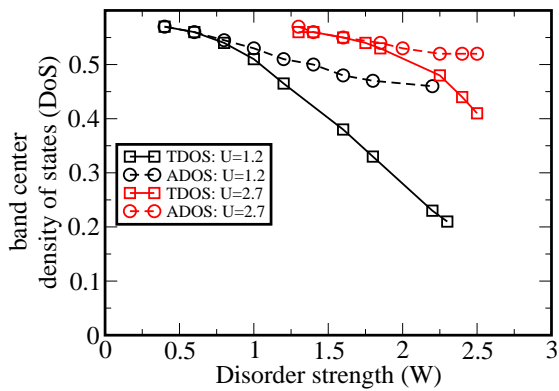}}
 \caption{Comparison of the band center ($\omega=0$) 
 value of the TDoS and the ADoS as a function of $W$.} 
  \label{fig:ADOS_TDOS_compare}
\end{figure}
To conclude this section, we compare the decay of the 
$\rho_{typ}(0)$ and the $\rho_{arith}(0)$ for the two regimes of interaction 
discussed above, namely, $U=1.2(< U_{c2})$ and $U=2.7(< U_{c2})$. 
From Fig.\ref{fig:ADOS_TDOS_compare}, $\rho_{typ}(0)$ appears to be monotonically vanishing as $W$ is increased while $\rho_{arith}(0)$ appears to saturate. 
If the metal-insulator transition encountered at large $W$ is continuous, as expected for small values of $U$, then these results suggest that the $\rho_{arith}(0)$  remains finite even in the insulating phase such that the Anderson-Mott insulator phase is gapless \cite{Byczuk2005}.
A true characterization of the phases would require numerical simulations very close to the 
metal-insulator transitions. The numerical calculations become very unstable as one approaches this 
limit and hence is beyond the scope of the current work.

\section{Conclusions}
We employed the dynamical mean field theory 
framework with a typical medium, to look into the 
interplay of disorder and strong correlations in the 
paramagnetic metallic phase of the 
particle-hole symmetric Anderson-Hubbard model using the 
local moment approach. 
Particularly, we explored the single particle dynamics by 
analyzing the disorder averaged self-energy and identified 
the existence of a universal `Kondo' scale within such a 
local theoretical framework that considers the strong correlation physics in presence of disorder scattering 
only due to 
deep trapped states. Additionally, we showed that this 
scale could be represented by the peak ($T_K^{peak}$)
in the distribution function of the 
Kondo scales. Moreover, the universal regime is shown to  
exist up to significantly high energies, 
although a strict Fermi liquid scattering dynamics 
holds true for $\omega\lesssim T_K^{peak}$. While 
such universal dynamics similar to that observed in the 
strong coupling limit of the conventional 
single-band Hubbard  model 
\cite{Logan_mott_ins, Zitko_Shastry} is anticipated 
in the low disorder regime \cite{disorder_Tneq0}, 
the same is surprising 
in the proximity of a Anderson-Mott transition, where 
the disorder is much stronger in comparison to the 
interaction. 
But then, such an observation highlights the incipient 
disorder renormalised Kondo screening of the 
local moments to be the dominant mechanism determining the 
low energy physics of the system. 

As mentioned before, within the local framework of 
the dynamical mean field theory in combination within the typical medium theory, the Anderson-Hubbard  model is mapped onto an ensemble of impurity problems, where the host for the impurities is determined by the typical density of states.  The tendency of an impurity site 
to form a local moment is governed by the impurity-host hybridization function that is determined by the typical impurity density of states. Thus the low energy physics will be determined by the peak of the distribution of the density of states and reinforced by this self consistency since all the sites see the same hybridization function. In this case, these sites are the ones with the lowest Kondo scale, which are at the peak of the distribution.  They are the ones closest to Mott character. The 
inhibition of the low energy hybridization function 
would be felt by all the impurities leading to a 
pronounced tendency towards forming local moments.

Since the disorder, especially near an Anderson localization transition, strongly suppresses the hybridization to the impurity, our observations highlight that 
in a disordered interacting system, Anderson and Mott 
mechanism of localization may not be disentangled. It is worth noting that the behavior of the local Kondo scales and the density of 
states are in agreement with the 
previous works as in Refs.~\onlinecite{Byczuk2005,Aguiar2009}.  In our work we perform a detailed investigation of the spectra, and find that the broad distribution 
of Kondo scales and the underlying universal scattering 
dynamics corroborate the physical picture 
of the emergence of the formation 
of local moments in the presence of metallic droplets, as 
proposed in Refs.~\onlinecite{Paalanen1988,Aguiar2009}. 
While the emergent local moments would tend towards a 
common Kondo scale, we speculate that the Kondo scales and 
hence the low energy physics associated with the metallic 
droplets could be inhomogeneous. 
These observations are 
particularly relevant for understanding the 
underlying mechanisms 
that lead to the breakdown of the metallic phase towards a 
Mott or Mott-Anderson localization transition. However, in 
order to assert 
the true nature of this spatial inhomogeneity we require 
to go beyond the local framework and incorporate non-local 
dynamical fluctuations. 

The local moment approach is an inherently non-perturbative impurity solver neither confined 
to low energies like the slave-boson approach nor to weak coupling like the iterated 
perturbation theory or modified perturbation theory approaches. 
While for non-disordered correlated systems this has clearly been 
demonstrated \cite{LMA_SIAM1,LMA_NRG_benchmark1}, our present calculations 
show that it does capture the 
strong correlation physics in accordance with the numerically exact NRG 
calculations for disordered correlated systems \cite{Byczuk2005}. 
With this set up established, one then asks the question as to what happens if 
we include short-range dynamical fluctuations 
due to disorder. Such directions 
within the framework of the typical medium dynamical cluster approximation 
\cite{Chinedu_TMDCA2014} are currently under our consideration.

\section*{Acknowledgments}
We would like to acknowledge fruitful discussions with Pinaki Majumdar and Subroto Mukerjee.  S.S. acknowledges the financial support from CSIR, India and JNCASR, India. 
This material is based upon work supported by the National Science Foundation award DMR-1237565 and 
by the EPSCoR Cooperative Agreement EPS-1003897 with additional support from the Louisiana Board of Regents. 
Supercomputer support is provided by the Louisiana Optical 
Network Initiative (LONI) and HPC@LSU. 

\appendix
\label{app1}
\section{Local moment approach (LMA)}
\label{app1}
\subsection{Starting point: unrestricted Hartree Fock}
In the following we will discuss some of the basic concepts of the zero 
temperature LMA formalism.  A key physical aspect of this method 
is the inclusion of low energy spin-flip excitations in the 
single-particle dynamics. This is facilitated at the inception by starting 
from the unrestricted Hartree Fock (UHF) state: local moments, $\bar{\mu}$, are 
introduced from the outset, to get a direct handle on the 
low energy spin-flip processes.
The solutions are built around simple symmetry broken static 
mean-field, UHF, states, containing two degenerate states 
$\bar{\mu}=\pm|\bar{\mu}|$, 
where, 
$|\bar{\mu}|=|\langle\hat{n}_{i\uparrow}-\hat{n}_{i\downarrow}\rangle|$, the 
average being over the UHF ground state. We label $A$ and $B$ for solutions 
$\bar{\mu}=+|\bar{\mu}|$ or $-|\bar{\mu}|$ respectively \cite{LMA_SIAM1}.
For an understanding of the 
formal details the reader is referred to \cite{LMA_SIAM1, 
LMA_asymm_SIAM, LMA_KIs, Raja_dyn_sca_pam, Eastwood1998, 
Dickens2001}.
Here, we briefly recap the main equations.
The single particle UHF Green's functions for the paramagnetic case, 
are given by,
\begin{align}
  &\mathcal{G}_{\uparrow}^{UHF}(\omega)
=\frac{1}{\omega^+-e_i+x-\Gamma(\omega)},
\label{eq:Gf_UHF(a)}\\
&\mathcal{G}_{\downarrow}^{UHF}(\omega)
=\frac{1}{\omega^+-e_i-x-\Gamma(\omega)},
\label{eq:Gf_UHF(b)}
\end{align}
where, $\Gamma(\omega)$ is the hybridization function for the impurity-host 
coupling, that, for the paramagnetic case, is spin
 independent; 
 $x=\frac{1}{2}|\bar{\mu}| U$ and $e_i=\epsilon_i+\frac{1}{2}Un$, where $n$ is the mean-field charge as described 
 in the following.  
The density of the single-particle excitations is given by, 
 $ \mathcal{D}_\sigma=-\frac{1}{\pi}\mathrm{Im}\mathcal{G}_\sigma^{UHF}(\omega)$,
where, $\sigma=\uparrow/\downarrow$.
The local moment, in general, would be given by,
$  |\tilde\mu|=
\int_{-\infty}^0d\omega\left(\mathcal{D}_\uparrow(\omega)-
\mathcal{D}_\downarrow(\omega)\right)$,
and has to be obtained {\it self-consistently}. When we are 
away from particle-hole symmetry then we also need the impurity occupancy 
to be given by,
 $ \tilde n=
\int_{-\infty}^0d\omega\left(\mathcal{D}_\uparrow(\omega)+
\mathcal{D}_\downarrow(\omega)\right)$.
For the pure mean-field UHF solution, we have, 
\begin{align}
&\tilde\mu=\bar{\mu} 
\label{eq:UHF_SC1}\\ 
&\tilde n=n, 
\label{eq:UHF_SC2}
\end{align}
to be solved self-consistently. So, if we now 
fix $x$ and $e_i$ (note that they are not the bare parameters of the 
Hamiltonian), then Eqs.~\eqref{eq:UHF_SC1} and~\eqref{eq:UHF_SC2} 
would provide the solution at one shot 
and accordingly, the bare parameters may be inferred as $U=2x/|\mu|$ and 
$\epsilon_i=e_i-Un/2$. However, if $U$ and $\epsilon_i$ are fixed then this 
has to be obtained by iterative cycling. 
The UHF solution is severely deficient 
(see \cite{LMA_SIAM1, 
glossop2002single, LMA_asymm_SIAM, LMA_gapless_fermi,LMA_KIs}), for not capturing 
the Fermi liquid picture. 
In any case, being a 
static approximation, one has to go beyond it to incorporate dynamics. 

\subsection{Inclusion of spin-flip scattering dynamics}

\begin{figure}[htp!]
\centerline{\includegraphics[clip=,scale=0.25]
{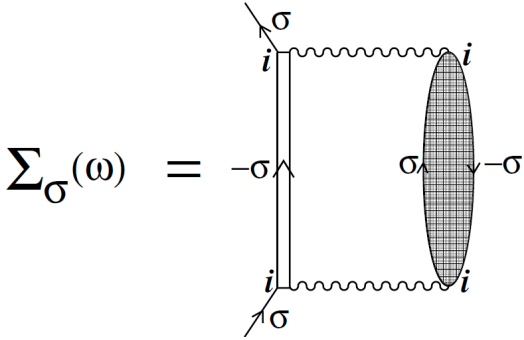}}
\caption
{{\bf Self-energy within the LMA for the single 
impurity Anderson model:} Diagrammatic 
representation of the dynamic self-energy, $\Sigma(\omega)$ 
retained within the LMA in practice. The diagrams are 
expressed in terms of the
polarization bubble, $\Pi^{\bar{\sigma}\sigma}(\omega)$.
Wavy line: interaction, $U$, double line: renormalized host/
medium propagator, hatched region: transverse spin polarization 
propagator bubble. See Eq.~\eqref{eq:SE_LMA} and the 
associated text.}
\label{fig:SE_LMA}
\end{figure}

 Within the LMA in practice, we approximate the dynamical part of the 
self-energy by the (nonperturbative) class of spin-flip diagrams shown
 in Fig.~\ref{fig:SE_LMA}.  Here, the 
bare propagators are that of UHF and therefore the inclusion of all these 
diagrams constitute the UHF+random phase approximation (RPA) scheme. 
We thus build a two self-energy 
description, as represented in Fig.~\ref{fig:SE_LMA} and mathematically 
represented as,
\begin{equation}
\Sigma_{\sigma}(\omega)=U^2\int_\infty^\infty\frac{d\omega^\prime}
{2\pi i}\mathcal{G}_{\bar{\sigma}}^{UHF}(\omega-\omega^\prime)
\Pi^{\bar{\sigma}\sigma}(\omega^\prime).
\label{eq:SE_LMA}
\end{equation}
$\Pi^{\bar{\sigma}\sigma}(\omega)$ is the transverse-spin polarization 
propagator (with $\bar{\sigma}=-\sigma$), which in the current RPA scheme employed is expressed as, 
$\Pi^{\bar{\sigma}\sigma}(\omega)=
\frac{^0\Pi^{\bar{\sigma}\sigma}}
{1-U^0\Pi^{\bar{\sigma}\sigma}}$. The bare polarization 
propagator, $^0\Pi^{\bar{\sigma}\sigma}(\omega)$ is expressed 
in terms of broken symmetry mean-field propagators as, 
$^0\Pi^{\bar{\sigma}\sigma}(\omega)=
\frac{i}{2\pi}\int_{-\infty}^{\infty}
   d\omega^\prime \mathcal{G}_\downarrow^{UHF}(\omega^\prime)
   \mathcal{G}_\uparrow^{UHF}(\omega^\prime-\omega)$.

The UHF propagators should result in a self-energy that satisfies the 
basic criteria for a Fermi liquid.
After some detailed algebra \cite{LMA_SIAM1}, we can then arrive at 
self-consistency equation for determining the exact 
local moment that satisfies such constraints, so that 
the two self-energy description may be written as,  
\begin{equation}
  \sum_\sigma\sigma\Sigma_\sigma(0; e_i,x)=|\tilde\mu(e_i,x)|U.
  \label{eq:symm_res}
\end{equation}
The above equation is known as the symmetry-restoration condition.
Finally, the single self-energy may be obtained as, 
\begin{align}
  \Sigma(\omega)=&\frac{1}{2}
\left[\tilde\Sigma_\uparrow(\omega)+\tilde\Sigma_\downarrow(\omega)\right]\nonumber \\
&+\frac{\frac{1}{2}\left(\tilde\Sigma_\uparrow(\omega)-
\tilde\Sigma_\downarrow(\omega)\right)^2}
{{g}^{-1}(\omega)-
\frac{1}{2}\left[\tilde\Sigma_\uparrow(\omega)+
\tilde\Sigma_\downarrow(\omega)\right]},
 \label{eq:single_SE}
\end{align}
where, 
\begin{equation}
  \tilde\Sigma_\sigma(\omega)=\frac{U}{2}(\tilde{n}-
  \sigma|\tilde\mu|)+\Sigma_\sigma(\omega),
\end{equation}
with $\sigma=\uparrow/\downarrow$ and the impurity Green's function,
$G_{imp}=\frac{1}{2}\sum_\sigma G_\sigma$, with 
$G_\sigma(\omega)=\left[{g}^{-1}(\omega)-
\tilde\Sigma_\sigma(\omega)\right]^{-1}$
and $g(\omega)=\frac{1}{\omega^+-\epsilon_i-\Gamma(\omega)}$.
Additionally, for p-h asymmetric situations \cite{glossop2002single}  
one also needs to satisfy the Luttinger's theorem given by, 
\begin{equation}
 I_L=\mathrm{Im} 
\int_{-\infty}^0\frac{d\omega}{\pi}\frac{\partial \Sigma(\omega)}
{\partial\omega}G_{imp}(\omega)=0.
\label{eq:LT}
\end{equation}
The self-consistent imposition of 
Eq.~\eqref{eq:symm_res} amounts to a self-consistency condition for the local 
moment $|\bar{\mu}|$ 
that enters Eqs.~\eqref{eq:Gf_UHF(a)},~\eqref{eq:Gf_UHF(b)}. A 
low energy spin-flip scale, $T_K$ is generated; this scale that manifests 
as a strong resonance in the imaginary part of the 
transverse spin polarization propagator, 
Im$\Pi^{\bar{\sigma}\sigma}(\omega)$, 
is proportional to the Kondo scale \cite{LMA_SIAM1, glossop2002single}. 
If the symmetry-restoration condition Eq~\eqref{eq:symm_res} 
is not satisfied then a 
spin-flip scale occurs at $T_K=0$ signaling the breakdown of a Fermi 
liquid.
     
A practical implementation of the LMA involves fixing 
$x=\frac{1}{2}U|\bar{\mu}|$ and $e_i$ 
\cite{LMA_SIAM1, LMA_asymm_SIAM, LMA_KIs, Raja_dyn_sca_pam,Raja2_2005}. 
The current nature of the problem, however, requires us
to fix the bare parameters of the single impurity Anderson model, 
namely, $U$ and $\epsilon_i$. However, 
we should also note that if $\epsilon_i$ is fixed instead of $e_i$ then, 
Eq.~\eqref{eq:symm_res} and Eq.~\eqref{eq:LT} would have to be solved 
self-consistently requiring several $\sim 20-30$ 
symmetry-restoration steps increasing the 
computation time enormously. Instead, if we fix $U$ and $e_i$ and tune 
$\bar{\mu}$ we can drastically reduce this requirement again ending up in 
solving 5-6 symmetry-restoration iterations, as in the fixed $x$, fixed $e_i$ algorithm. The scheme is described as following:
\begin{enumerate}
\item{We start with an initial guess local moment, 
$\bar{\mu}$ with which we 
calculate $\mathcal{G}_{\uparrow/\downarrow}$ and subsequently, $\tilde\mu$ 
from UHF spectral functions.}
\item{The calculation of $^0\Pi^{\bar{\sigma}\sigma}$ and $\Pi^{\bar{\sigma}\sigma}$, and, 
$\Sigma_{\uparrow/\downarrow}$ follows.}
\item{Eq.~\eqref{eq:symm_res} is checked and steps (1), (2), (3) are repeated 
until a convergence of $~10^{-6}$ or lower 
is achieved. 
With this step it can be realized that the 
entire process involves calculations of coupled equations for finding the 
root of Eq.~\eqref{eq:symm_res}, for which one therefore has to provide a 
judicious guess to reach the solution correctly and efficiently.}
 \item{Finally, with proper guesses for the underlying self-consistency 
equations a converged $\Sigma(\omega)$ is obtained. With this 
self-energy, we can now satisfy 
Eq.~\eqref{eq:LT} by tuning $\epsilon_i$.}
\end{enumerate}
In particular to the problem treated in this paper, 
we also had to take care of the computation time required to be able 
to sample sufficient number of disorder realizations. We achieved this 
by bringing in some additional {\it schemes} which would be discussed 
in detail in the following section.

\section{Numerical implementation of TMT-DMFT}
\label{app2}
In this section we provide
technical details of our implementation of the LMA within the TMT-DMFT 
framework. For the sake of completeness we also outline the steps involved 
in the TMT-DMFT implementation.
As outlined in the previous section, 
employing the LMA with the bare parameters $U$ and $\epsilon_i$,
would require a lot of computational time.
This results from the fact that, away
from p-h symmetry the impurity parameter $e_i$ that acts like a
{\it pseudo chemical potential} and explicitly enters the UHF 
Green's functions via
Eqs.~\eqref{eq:Gf_UHF(a)},~\eqref{eq:Gf_UHF(b)}, would have to be
tuned so that the symmetry-restoration 
(Eq.~\eqref{eq:symm_res}) and the Luttinger's theorem
(Eq.~\eqref{eq:LT}) are self-consistently satisfied. Recall that this
would result in repeating the symmetry restoration 
(Eq.~\eqref{eq:symm_res}) step described in several times.
Instead, the impurity self-energy may be obtained at a much cheaper effort
if the bare parameters $U$ and the impurity parameter $e_i$ is fixed.
In that case, once the symmetry restored impurity self-energy and Green's
functions are obtained, one can tune the $\epsilon_i$ such that the
the Luttinger's theorem (Eq.~\eqref{eq:LT}) is satisfied. This can be done
without having to repeat the impurity self-energy calculation.
However, in the current problem, the $\epsilon_i$ is a random
quantity following a particular distribution. So, in order to resort
to the fixed $U$, fixed $e_i$ scheme discussed in the earlier section 
we have to first build a
{\it database} for the respective $(e_i,\epsilon_i)$ pair with the
given hybridization.
In other words, before going to the actual calculation we do the following:

{\bf Step 1:}
\begin{enumerate}
  \item{Given a hybridization function, 
  $\Gamma(\omega)$ we start from
the particle-hole symmetric limit with $e_i=0$ and $\epsilon_i=-U/2$, for
which the Luttinger's theorem (Eq.~\eqref{eq:LT}) is naturally satisfied. Note that in the main text, $\Gamma(\omega)$, 
has been denoted as $\Gamma_{typ}(\omega)$. So,
in this step the LMA solver is provided with (a) $\Gamma(\omega)$, (b) $U$,
(c) $e_i=0$.}
\item{We now increment the $e_i$ by a small step, say 0.02\footnote{This is
optimized by experience to minimize the number of steps or
$(e_i,\epsilon_i)$ pairs required to obtain a {\it good} database.}.
So, in this step the LMA solver is provided with (a) $\Gamma(\omega)$,
(b) $U$, (c) $e_i=0.02$. Accordingly, the $\epsilon_i$ is derived by
satisfying the Luttinger's theorem (Eq.~\eqref{eq:LT}) and an
$(e_i,\epsilon_i)$ pair for the given $\Gamma(\omega)$ is generated.}
\item{The above step (2) is continued until the $\epsilon_i$ obtained
overshoots the limit set by the disorder strength, $W$. Note that,
$\epsilon_i=-U/2+V_i$, where $V_i$ is a random number between
$-W\leq V_i\leq W$.}
\item{For the actual random configuration, $V_i$, and therefore, the
$\epsilon_i$, we now interpolate the corresponding $e_i$ from the database
and compute the local self-energy, $\Sigma_i$. Finally, we construct the local
Green's function, $G_i(\omega, V_i)$, using the equation, 
$G_i(\omega, V_i)=\left([\mathcal{G}]^{-1}(\omega)
-\Sigma_i(\omega)-\epsilon_i\right)^{-1}$, where, 
$\mathcal{G}(\omega)=\frac{1}{\omega^+-\Gamma(\omega)}$.
This would now be used to construct $\rho_{typ}(\omega)$.}
\end{enumerate}

{\bf Step 2:}

The output of the {\bf Step 1} comprises $N$ local impurity self-energies, 
$\Sigma_{i}(\omega)$ that gives us $N$ local impurity Green's function, 
$G_{i}(\omega)$. With the local spectral functions, 
$\rho_{i}(\omega)=-\frac{1}{\pi}\mathrm{Im}G_{i}(\omega)$, we construct the 
disorder averaged DoS, using geometric averaging:
\begin{equation}
  \rho_{typ}(\omega)=\exp\int dV_iP(V_i)\ln\rho_i(\omega)
  \label{eq:rho_typ1}
\end{equation} 
Using Eq.~\eqref{eq:rho_typ1} we can now construct the {\it typical} Green's 
function, $G_{typ}(\omega)$, from the Hilbert transform of $\rho_{typ}$:
\begin{equation}
  G_{typ}(\omega)=\int\frac{\rho_{typ}(\omega^\prime)
d\omega^\prime}{\omega-\omega^\prime}.
\label{eq:Gtyp}
\end{equation} 

{\bf Step 3:}
 We define the coarse-grained lattice Green's function 
 as $\overline{G}(\omega)$, given by,
\begin{equation}
   \overline{G}(\omega)=\int\;\frac{\rho_0(\epsilon)\;d\epsilon}
   {\left[G_{typ}(\omega)\right]^{-1}+\Gamma(\omega)-\epsilon},
\label{eq:G_latt}
\end{equation}
where $\rho_0(\epsilon)$ refers to the bare density of states, that in the 
current problem is that of the 3-dimensional cubic lattice.

{\bf Step 4:}
The new hybridization may be obtained as,
\begin{equation}
  \Gamma(\omega)_{new}=
\Gamma_{old}+\zeta\left[(G_{typ})^{-1}-(\overline{G})^{-1}\right],
\label{eq:new_gamma}
\end{equation}
where, $\zeta$ is a mixing parameter typically set to a value of 0.5.
With $\Gamma_{new}(\omega)$ we can go back to {\bf Step 1} and continue until 
-Im$\int|\left(\Gamma_{new}(\omega)-\Gamma_{old}(\omega)\right)|d\omega$ 
converges within some tolerance, which in our 
implementation is chosen to be $\sim 10^{-3}$.

Note that in order to look into the scattering dynamics, 
we 
calculate the arithmetic average of the local density of 
states, $\rho_i(\omega)$. As described in the main 
text, this is given by, 
$\rho_{arith}=\int dV_iP(V_i)\rho_i(\omega)$ and it 
represents the average density of states (ADoS) of the 
lattice. From the ADoS, we can then calculate 
the arithmetic average of the local Green's function, 
$\langle G(\omega) \rangle_{arith}$, using the, 
Hilbert transform relation, 
$\langle G(\omega)\rangle_{arith}=
\int\frac{\rho_{arith}(\omega^\prime)
d\omega^\prime}{\omega-\omega^\prime}$. 
The 
disorder-averaged self-energy, $\Sigma(\omega)$, that 
represents the scattering dynamics is then calculated as, 
$\Sigma_{ave}(\omega)
=\mathcal{G}(\omega)^{-1}-
\langle G(\omega)\rangle_{arith}^{-1}$.

\section*{References}
\bibliography{paper_Nc1}

\end{document}